\documentclass[conference]{IEEEtran}
\IEEEoverridecommandlockouts
\usepackage{cite}
\usepackage{amsmath,amssymb,amsfonts}
\usepackage{algorithmicx}
\usepackage{algorithm}
\usepackage{algpseudocode}

\usepackage{graphicx}
\usepackage{textcomp}
\usepackage{xcolor}
\usepackage{soul}
\usepackage{booktabs}
\def\BibTeX{{\rm B\kern-.05em{\sc i\kern-.025em b}\kern-.08em
    T\kern-.1667em\lower.7ex\hbox{E}\kern-.125emX}}
\begin{document}

\title{Enabling Intelligent Vehicular Networks Through Distributed Learning in the Non-Terrestrial Networks 6G Vision\thanks{This work was partially supported by the European Union under the Italian National Recovery and Resilience Plan (NRRP) of NextGenerationEU, partnership on “Telecommunications of the Future” (PE00000001 - program “RESTART”)}}
\author{\IEEEauthorblockN{David~Naseh, Swapnil Sadashiv Shinde, and Daniele~Tarchi}
\IEEEauthorblockA{\textit{Department of Electrical, Electronic and Information Engineering “Guglielmo Marconi”}\\
\textit{Alma Mater Studiorum - Università di Bologna}\\
40136 Bologna, Italy\\
\{david.naseh2,swapnil.shinde2,daniele.tarchi\}@unibo.it}
}

\maketitle

\begin{abstract}
The forthcoming 6G-enabled Intelligent Transportation System (ITS) is set to redefine conventional transportation networks with advanced intelligent services and applications. These technologies, including edge computing, Machine Learning (ML), and network softwarization, pose stringent requirements for latency, energy efficiency, and user data security. Distributed Learning (DL), such as Federated Learning (FL), is essential to meet these demands by distributing the learning process at the network edge. However, traditional FL approaches often require substantial resources for satisfactory learning performance. In contrast, Transfer Learning (TL) and Split Learning (SL) have shown effectiveness in enhancing learning efficiency in resource-constrained wireless scenarios like ITS. Non-terrestrial Networks (NTNs) have recently acquired a central place in the 6G vision, especially for boosting the coverage, capacity, and resilience of traditional terrestrial facilities. Air-based NTN layers, such as High Altitude Platforms (HAPs), can have added advantages in terms of reduced transmission distances and flexible deployments and thus can be exploited to enable intelligent solutions for latency-critical vehicular scenarios. With this motivation, in this work, we introduce the concept of Federated Split Transfer Learning (FSTL) in joint air-ground networks for resource-constrained vehicular scenarios. Simulations carried out in vehicular scenarios validate the efficacy of FSTL on HAPs in NTN, demonstrating significant improvements in addressing the demands of ITS applications.
\end{abstract}


\section{Introduction}
With the 6G vision set to shape the 2030s society into a more advanced, digitized, fully connected, and intelligent world, transportation networks are also going through the transition of converging into Intelligent Transportation Systems (ITS) \cite{8926369}. The field of ITS is expanding quickly and seeks to enhance the safety and efficiency of traditional transportation systems, as well as sustainability, by leveraging technologies such as Machine Learning (ML), the Internet of Things (IoT), and advanced modes of communication. With the presence of IoT subsystems, a large amount of data can be generated over time in the various parts of the ITS system \cite{8767247}. These data can be effectively utilized to provide intelligent solutions. However, a significant challenge in ITS lies in acquiring and analyzing data from various sources, including vehicles, sensors, and traffic cameras. Lately, a variety of Distributed Learning (DL) techniques have emerged to appropriately address these challenges. Among others, Federated Learning (FL) has shown great promise to train the ML models effectively in a distributed manner \cite{9650754}.
%firstly, by enabling the distributed collection of data from diverse sources, ensuring a wide range of data is available for training ML models. Additionally, DL facilitates the distributed training and analysis of these models, leveraging parallel processing across multiple nodes or devices. Moreover, by utilizing DL, ITS can overcome data analysis challenges, train accurate ML models, and achieve efficient predictions for traffic patterns, collision avoidance, and driver assistance while maximizing the utilization of available data and improving scalability \cite{shinde2023joint}.

%A significant obstacle for ITS is the acquisition and analysis of this data from diverse sources, such as vehicles, sensors, and traffic cameras. Nevertheless, such data can be utilized to train advanced ML (ML) models that can predict traffic patterns, avoid collisions, and assist drivers.

 %Another challenge is the sensitivity of the data, such as data concerning the location of vehicles or the identities of drivers. The protection of data privacy is crucial during the collection, storage, and utilization of data. In the realm of ITS, DL techniques such as federated learning (FL), split learning (SL), and transfer learning (TL) can be applied to overcome these challenges associated with training ML models.

FL is a DL paradigm that facilitates the shared training of a model without the exchange of individual data among multiple devices. This technique enables each device to train a local model on its data and then update a central server with the model's progress. The central server then aggregates the updates from all devices and applies them to the shared model. The central server subsequently aggregates the updates from every device and applies them to the global shared model. One of the main drawbacks of FL is that each client needs to train the entire ML model, which is unaffordable to clients with limited resources, such as those found in ITS, especially when the ML model under training is a Deep Neural Network (DNN) \cite{9650754}. Furthermore, as a result of the server and clients' complete access to the local and global model parameters during training, in recent times, several new privacy concerns have also been raised from both the client and server sides, e.g., poisoning, attacks, and model inversions.

%To address these issues, SL is introduced.
Split Learning (SL) is another DL approach that allows complex ML models to be trained by dividing them into two parts, each trained on a client or a server using local data from distributed clients \cite{9652119}. In contrast to running an entire network, as in FL, only a portion of the network is assigned to train on the client side, significantly reducing the processing load for DL computation on devices with limited resources. Additionally, due to the split networks, the only communication between split parts is the activation of the cut layer, consequently, neither a client nor a server can access the other's model. Therefore, if integrated with FL, SL can help reduce the overall cost of FL training, increase the number of participating nodes, and enable FL to train more advanced DNNs. In recent times, some researchers have demonstrated these advantages by proposing SL-inspired FL frameworks \cite{9923620}.  

Transfer Learning (TL), from the meta-learning family, is another learning tool that has been extensively explored in ML research, especially for increasing the efficiency of the training processes \cite{Vilalta2022}. The key fundamental of TL is knowledge transfer (KT), which is to utilize and transfer knowledge and experience gained from similar source tasks in the past to facilitate the learning of new related target tasks. Thus, TL approaches can increase the convergence rate, minimize reliance on labelled data, and improve the robustness of machine learning techniques in different vehicular settings. With these advantages, TL can complement the FL process, especially in resource-constrained ITS scenarios with stringent application requirements. Recently, some authors have tried to merge the advantages of TL in the FL setting by proposing TL-inspired FL models \cite{10005134}.

In the 6G vision, Non-Terrestrial Networks (NTNs) have acquired a central position, mainly to provide global coverage and capacity boosts. Different NTN platforms can help Vehicular Users (VUs) to enable intelligent solutions. In particular, High-Altitude Platforms (HAPs), with their reduced transmission distances, higher coverage, and easy and flexible deployments, can complement terrestrial vehicular settings to enable efficient DL solutions \cite{shinde2023joint}.  

With this motivation in mind, in this work, we propose a novel DL methodology titled Federated Split Transfer Learning (FSTL) to effectively train ML models for NTN-based resource-constrained ITS scenarios. Then, we evaluate the proposed method in a vehicular scenario with AlexNet on the MNIST dataset, and the results demonstrate better performance compared to traditional FL methods in terms of convergence rate, training accuracy, and overall latency.

%The rest of this paper is organized as follows: Section 2 presents the proposed technique. Section 3 evaluates the proposed technique with simulations. Section 4 discusses the advantages and challenges of the proposed technique and future work. Finally, we conclude the paper.

\section{Federated Split Transfer Learning}
\subsection{Introduction to the Elements}
%First, we will start by formally describing the process of FL. 
The considered vehicular scenario includes a set of $N$ distributed VUs \(\mathcal{V}=\{v_1,\cdots,v_i,\cdots,v_N\}\) having their own labeled dataset  \(\mathcal{D}_i=\{(\textbf{x}_k,y_k)\}\), for {$1\le k\le K_i$}, with $K_i$ data samples. Here, $\textbf{x}_k \in \mathcal{R}^n$ is the $n$ dimensional feature vector associated with the $k$th data sample, while $y_k$ is the corresponding label. The VUs aim to solve a generic learning task $p$ through a collaborative FL framework. The objective is to learn a global model \(W_p\) that minimizes a given loss function \(L^p\) across all participants without explicitly sharing the raw data. 

In each FL round \(t\), the participant \(v_i\) updates its local model parameters by performing a gradient descent step on its local loss function \(L^{p}_{i}(\mathcal{D}_i, W_t)\), where \(W_t\) represents the global model at round \(t\) based upon the parameter updates received by the server in the previous round. The updated local parameters are then sent to a central server, which aggregates the model updates across participants using a weighted averaging scheme (e.g., FedAvg) to obtain the global model defined as $W_{t+1}= \frac{1}{N}\sum_{\forall i \in N} W_{i,t}^{p}$, where $W_{i, t}^{p}$ is the local model parameters from VU $i$. The FL process iterates till it achieves predefined stopping criteria, e.g., number of iterations, and loss function convergence value.

The above-mentioned FL approach offers numerous advantages in the realm of ITS. First, it enables collaborative learning from diverse data sources 
% distributed across different entities such as vehicles, infrastructure, and cloud servers,
without the need for data centralization. This preserves data privacy, security, and regulatory compliance, which are of utmost importance in ITS. Second, FL allows for continuous model improvement by leveraging real-time data, leading to improved prediction and decision-making in ITS applications. Furthermore, it reduces communication overheads since only model parameters are exchanged between participants rather than raw data.
% This is particularly crucial in resource-constrained environments such as vehicles with limited bandwidth. 
%Overall, FL empowers ITS systems to collectively learn and adapt to dynamic traffic conditions while respecting data privacy and fostering cooperation among multiple stakeholders.

% However, in some ITS scenarios, there might be restrictions on sharing even the model updates due to privacy concerns, limited communication capabilities, or computation constraints. In such cases, SL emerges as a viable approach.

Though FL has several advantages in terms of distributed model training, data privacy, and data transmission costs, it is often the case that FL requires a large number of iterations to achieve satisfactory performance. Each FL iteration puts an additional burden on the resource-constrained ITS nodes in terms of latency, energy, etc. Also, with the repeated communication rounds between VUs and server nodes, several new privacy threats emerged where third-party intruders could influence the FL process and even breach the FL data security over time. Therefore, it is quite challenging to fully train and communicate  complex ML models in traditional centralized FL environments. 

SL is an ML paradigm that allows the efficient training of complex ML models through various model split approaches. For example, in the considered IoV scenario, the model $W^p$ can be split into two parts: (i) VUs' split model (e.g., vehicles) \(S\), and (ii) a server split model ("merge" part) \(M\) that operates on the server side (e.g., Road Side Units (RSUs)). The $i$th VU split part processes the raw data locally and transmits only the intermediate representations \(H_i = S(\mathcal{D}_i)\) to the server part, which performs further calculations and updates the global model \(W^p\) through collaborative learning procedures. With this approach, VUs can effectively train complex ML models and communicate only the partial model parameters. This can effectively solve the two main issues of the traditional FL process discussed previously, i.e., training cost and privacy issues.
 
 % This approach addresses the challenges of data privacy and communication limitations in ITS, allowing for collaborative learning while minimizing data transmission and preserving privacy. SL is particularly useful when the raw data cannot be easily shared due to privacy regulations or when bandwidth constraints prevent direct model updates. SL procedure is as follows.

% Let \(\mathcal{R}_i\) represent the raw data held by the \(i\)-th participant in a distributed system. The objective of Split Learning is to learn a global model \(W\) by dividing the model architecture into two parts: a "split" part denoted by \(S\) that operates on the client side (e.g., vehicles), and a "merge" part denoted by \(M\) that operates on the server side. The split part processes the raw data locally and produces intermediate representations \(H_i = S(\mathcal{R}_i)\). These intermediate representations are then transmitted to the merge part, where they are merged with other participants' representations and used to update the global model \(W\) through collaborative learning procedures.

Although SL addresses certain challenges in ITS, it introduces disadvantages such as the limited expressive power of the split part, the risk of model performance degradation due to the separation of computations, and high delays because of its serial training instead of parallel training as in FL. To overcome these limitations, a combination of FL and SL, referred to as SplitFed or Federated Split Learning (FSL), is employed \cite{9652119}. SplitFed leverages the collaborative learning and privacy-preserving properties of FL while benefiting from the improved local data processing capabilities of SL. The procedure can be defined as follows. The participant \(v_i\) processes \(\mathcal{D}_i\) locally using the split part of the model to obtain intermediate representations \(H_i\). These representations are subsequently shared with a central server, where they are merged with other participants' intermediate representations, i.e. federated averaging. The merged representations are used to update the global model \(W^p\) through FL procedures, fostering collaboration and model improvement across participants.

The performance of the proposed FSL approach can be further improved through the integration of TL solutions. In the case of TL, through the use of a pre-trained neural network model, initial knowledge can be transferred to the local devices, providing a valuable head start for the training process. Thus, in a considered solution, each VU receives a pre-trained ML model $W^{p'}$ associated with task $p'$ belonging to the same family as $p$. This approach can speed up the convergence of the model $W^p$ and reduce the amount of time and computational resources needed for local training. Moreover, TL allows devices with limited computational capabilities to participate in the FSL process, thus enhancing the training performance of the FSL model. In the next section, we will go into detail on the proposed FSTL methodology.

\subsection{Construction of FSTL Platform}
In the approach considered, a pre-trained neural network model $W^{p'}$ with \(L\) layers. Next, $W^{p'}$ is split at a specific layer $k$, called the cut (or splitting) layer, such that \(1 < k < L\), with the first part $W^{p'}_{VU}$ serving as the VU split model and the rest $W^{p'}_S$ as the SL server-side model. During the training process, each $i$th VU uses the vehicular data $D_i$ and the VU-side TL model $W^{p'}_{VU}$ to produce intermediate representations $H_i$ that are then transmitted to the SL server-side model ($W^{p'}_{S}$). The SL server-side model performs further computations using these intermediate representations and updates the global model parameters.

% %The VU split model ($W^{p'}_{VU}$) can be defined as:
% \[
% VU(x) = f_{VU}(x) = M^{(1:k)}(x)
% \]
% where \(x\) represents the input data, \(f_{VU}\) is the VU split function, and \(M^{(1:k)}\) denotes the sub-model of \(M\) containing layers from 1 to \(k\).

% The SL server-side model (\(SL\)) can be defined as:
% \[
% SL(h) = f_{SL}(h) = M^{(k+1:L)}(h)
% \]
% where \(h\) represents the intermediate representations produced by the VU split model (\(VU\)), and \(f_{SL}\) is the SL server-side function. \(M^{(k+1:L)}\) denotes the sub-model of \(M\) containing layers from \(k+1\) to \(L\).

\begin{figure}[tbp]
\centering
\includegraphics[width=\columnwidth]{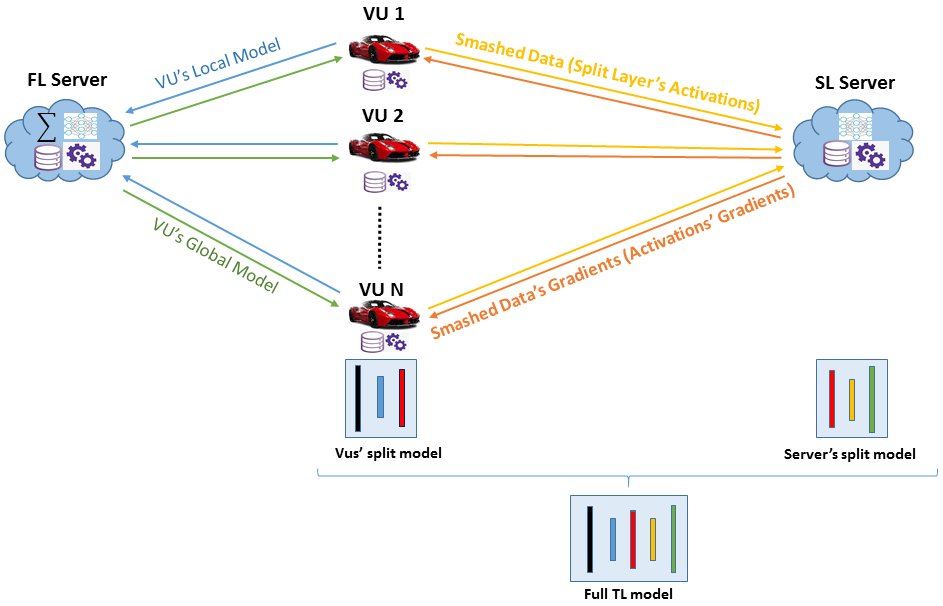}
\caption{Overall structure of FSTL}
\label{Fig:FSTL} 
\end{figure}

By introducing TL to FSL, we can leverage the advantages of pre-trained models in the VU split model ($W^{p'}_{VU}$). The pre-trained layers up to the splitting layer \(k\) capture general patterns and features, while the remaining layers in the SL server-side model ($W^{p'}_{S}$) enable collaborative learning and model improvement across distributed VUs. As observed in Fig. \ref{Fig:FSTL}, unlike SL, all clients communicate with the SL and FL servers simultaneously while conducting their computations in parallel, so a higher convergence rate is expected. This integration also allows efficient knowledge transfer, enhanced model performance, and faster convergence in the vehicular scenario while addressing resource constraints and privacy concerns.

%The proposed framework brings several advantages to the table. Firstly, it ensures data privacy and security by allowing data to remain on individual devices during training. This addresses concerns related to sensitive information and legal regulations surrounding data sharing. Secondly, the integration of SL reduces communication overhead, making the framework more suitable for resource-constrained devices and networks. Thirdly, by incorporating Transfer Learning, the framework enables faster convergence and improved performance by leveraging pre-trained models and initial knowledge transfer. Overall, the proposed approach combines the strengths of FL, SL, and TL to create an efficient and privacy-preserving ITS framework that can effectively handle data constraints, optimize model training, and enhance the intelligence of transportation systems.
%FSTL combines the main advantages of FL, TL, and SL, which include distributed parallel client processing, transfer of knowledge, and network splitting into two sub-networks during training for improved privacy, with the ability of TL to transfer knowledge from pre-trained Deep Neural Networks (DNN) to speed up training.

%A client might be a Vehicular User (VU), or also a hospital, or an Internet of Medical Things (IoMT) device with limited computing power and resources, while the central server could be a cloud server or a researcher with a powerful supercomputer, or even a High-Altitude Platform (HAP). To run Federated Averaging on the VU's local updates, the FL Server is presented. The VU global model is also synchronized by the FL server in each network training cycle. The aggregation performed by the FL Server, which mostly computes federated averaging, is affordable. The FL Server can therefore be deployed inside the local edge boundaries. On the other hand, the SL Server can carry out FL Server operations if all operations are implemented at the FL Server through encrypted data aggregation.

\subsection{Training Process in FSTL}

If we assume $W^{p'}_{VU}=\{\theta_i\}$ and $W^{p'}_{S}=\{\theta_s\}$ where $\theta_i$ and $\theta_s$ are model parameters for the VU-side and server-side subnetworks, then the training procedure of the FSTL can be described as follows:

\begin{enumerate}
    \item \textbf{Initialization}
    \begin{enumerate}
        \item Initialize the global model parameters $\theta$ with pre-trained weights from a TL model.
        \item Select a cut layer index \(k\) in the pre-trained model.
    \end{enumerate}
\item \textbf{Iterative Training.} For each training iteration \(t = 1, 2, \ldots, T\)
\begin{enumerate}
    \item   Distribute the global model parameters $\theta$ to the participating VUs.
     \item   Each VU \(i\) processes its local vehicular data \(\mathcal{D}_i\) using the VU split model $W^{p'}_{VU}$ to obtain intermediate representations:
       \[H_i = W^{p'}_{VU}(\mathcal{D}_i)\]\
     \item   VUs share their intermediate representations \(H_i\) with a central server.
     \item   The central server merges the intermediate representations \(H_i\) with a merge function \(Merge(\cdot)\):
       \[H_{\text{merged}} = Merge(H_1, H_2, \ldots, H_N)\]
     \item   The merged representations \(H_{\text{merged}}\) are used to update the global model parameters $\theta$ through FL procedures:
       $$\theta_i' = W^{p'}_{VU}(H_{merged}, \theta)$$
     \item   Update the global model parameters:
       \[\theta = \theta'\]
     \item   Each VU \(i\) updates its SL server-side model parameters \(W_i\) using the updated global model parameters:
       \[\theta_i' = W^{p'}_{S}(H_i, \theta)\]
     \item   Update the VU \(i\)'s model parameters:
       \[\theta_i = \theta_i'\]
\end{enumerate}
\end{enumerate}
The iterative training process is repeated until convergence or a predefined stopping criterion is met.

In this process, the VU split model $W^{p'}_{VU}$ performs local processing on vehicular data to generate intermediate representations. These representations are then shared and merged at the central server, allowing collaborative learning and model improvement. The merged representations are used to update the global model parameters \(\theta\) through FL procedures. Each VU then uses the updated global model parameters to update their model parameters \(\theta_i\). This iterative process enables the collective learning and adaptation of the FSTL model across distributed vehicular units while preserving privacy, addressing resource constraints, and leveraging TL to enhance model performance in the vehicular scenario.

% All VUs simultaneously forward propagate their local models and send the split data to the SL Server. Next, in parallel, the SL Server conducts forward propagation and backpropagation on its server model using the split data from each VU. The gradients of the split data are then sent to the relevant VU for backpropagation. Following that, the server updates its model by performing a federated weighted averaging of the gradients it computed using the split data from each VU throughout the backpropagation process. Each VU performs backpropagation on its local model and calculates its gradients at its end after receiving the gradients of its split data. To make these gradients more private and communicate them to the FL server, a Differential Privacy (DP) method can be utilized. All participating VUs get the federated average of the VU local updates from the FL Server.

\subsection{Latency Analysis}
Here we perform the total latency analysis modeled as the sum of the computation and communication time, for different DL methods, including FL, SL, FSL, and FSTL, with uniform data distribution. Assume that $d$ and $h$ are the total data size and the smashed layer's size, respectively, $R$ is the data rate between VUs and server platform based upon the communication medium, $T$ is the training time required to train full DNN model while $T'$ is the training time in FSTL where a pre-trained TL model is used, $T_{FedAvg}$ and $T_{Merge}$ are the time required to perform the full model and smashed parameters aggregation, respectively, $p$ is the total number of full model parameters, and $r$ is the ratio of the VU-side submodel's size to the full model’s size available in split-based models, i.e., SL, FSL and FSTL. The summary of the results is presented in Table \ref{tab:my_label}. Factor 2 in the terms, such as $2pr$, $\frac{2dh}{N}$, etc., is due to the download and upload of the VU-side model updates before and after training. As shown in this table, in case of a large number of clients, SL can become inefficient because the total time is proportional to $N$, mainly due to the serial training operation, whereas other methods perform the training in parallel. Furthermore, we can see that as $N$ increases, the total time cost increases in this order: $FSTL<FSL<FL<SL$. To be more specific, FSTL is faster than FSL because it uses a pre-trained network ($T'<T$), and FSL has less latency than FL because it aggregates fewer parameters ($T_{Merge}<T_{FedAvg}$). %These evaFinally, we observe this in the Simulations and Performance Evaluations section (Section IV). 

\begin{table*}[tbp]
%\scriptsize
    \centering
        \caption{Latency analysis of four DL methods per round}
    \label{tab:my_label}
\resizebox{\textwidth}{!}{  
\begin{tabular}{p{2.5cm}|p{3.5cm}|p{3cm}|p{3cm}|p{3cm}|p{3cm}}
\hline
\toprule
\centering
Learning Method & Training/Aggregation Time &  Communications per VU & Total communications & Total communication time & Total latency
\\
%\hline
\midrule
\centering
FL
&
\centering
$T+T_{FedAvg}$
&
\centering
$2p$
&
\centering
$2Np$
&
\centering
$\frac{2p}{R}$
&
$T+T_{FedAvg}+\frac{2p}{R}$
\\
\hline
\centering
SL
&
\centering
$T$
&
\centering
$\frac{2dh}{N}+2pr$
&
\centering
$2dh+2Npr$
&
\centering
$\frac{2dh}{R}+\frac{2Npr}{R}$
&
$T+\frac{2dh}{R}+\frac{2Npr}{R}$
\\
\hline
\centering
FSL
&
\centering
$T+T_{Merge}$
&
\centering
$\frac{2dh}{N}+2pr$
&
\centering
$2dh+2Npr$
&
\centering
$\frac{2dh}{NR}+\frac{2pr}{R}$
&
$T+T_{Merge}+\frac{2dh}{NR}+\frac{2pr}{R}$
\\
\hline
\centering
FSTL
&
\centering
$T'+T_{Merge}$
&
\centering
$\frac{2dh}{N}+2pr$
&
\centering
$2dh+2Npr$
&
\centering
$\frac{2dh}{NR}+\frac{2pr}{R}$
&
$T'+T_{Merge}+\frac{2dh}{NR}+\frac{2pr}{R}$
 \\
 %\hline
 \bottomrule
\end{tabular}
}
\end{table*}

\section{FSTL in NTN-based Vehicular Scenario}

NTNs have emerged as a promising solution to extend connectivity and support advanced applications in challenging environments \cite{shinde2023joint}. Vehicular scenarios, characterized by high mobility, intermittent connectivity, and dynamic data distribution, pose significant hurdles to traditional ML techniques. By integrating the capabilities of HAPs with FSTL, we harness the potential of NTNs to enable efficient and secure training and inference for VUs in vehicular networks.

In our proposed scenario, HAPs serve as both SL and FL Servers for VUs. SL enables the VUs to keep the raw data on their local devices while offloading computationally demanding tasks to the HAPs, such as model training. This approach preserves privacy and reduces the communication burden, as only model updates are transmitted between the VUs and the HAPs. The FL paradigm also enables collaborative model training across multiple VUs, promoting knowledge sharing and adaptation to diverse vehicular environments.

The integration of FSTL with HAPs in the NTN context opens up new possibilities for vehicular scenarios. Firstly, the high-altitude placement of HAPs ensures broader coverage and reduced interference, enabling seamless connectivity for VUs even in remote areas or areas with limited terrestrial infrastructure. Secondly, the FL architecture fosters collective intelligence among VUs, leading to improved models that can adapt to varying conditions and scenarios encountered by different vehicles. By leveraging NTNs and FSTL on HAPs, our proposed approach paves the way for intelligent vehicular networks that offer enhanced connectivity, privacy preservation, and improved decision-making capabilities.

Here, FSTL is proposed in vehicular scenarios to enable collaborative learning across multiple vehicles while preserving data privacy. The procedure involves a central server (SL Server, which here is a HAP) and multiple VUs exchanging model updates and gradients. 

\begin{algorithm}[H]
\caption{FSTL iterative algorithm}
\label{alg:VIt}
\small
\begin{algorithmic}[1]
\Require{$ N, \theta_i(0), \theta_s(0), \sigma, \alpha, \eta$} 
\Ensure{$\theta_i, \theta_s$}
\Statex
\State  Initialize server $\theta_s=\theta_s(0)$
\For {$1\le i\le N$}
\State  Initialize vehicular user $\theta_i=\theta_i(0)$
\State  Forward propagate on the $VU_i$ model using local data
\State  Send $H_i$ to HAP
\State  Forward propagate on the server model using $H_i$
\State  Back propagate ${\nabla\theta_s}$ on HAP
\State  Back propagate on the $VU_i$ model using ${\nabla\theta_s}$
\State  Update $VU_i$ local model using an optimizer (e.g., SGD): 
\begin{center}
    ${{\theta_i} \leftarrow {\theta_i} - {\eta}\cdot{\nabla\theta_i}}$ 
\end{center}
\State  Aggregate (federated average) on HAP: 
\begin{center}
   ${G_{avg} = \frac{1}{N} \cdot \sum(w_i \cdot G_i)}$ 
\end{center}
\State  Update HAP model using an optimizer (e.g. SGD): 
\begin{center}
    ${{\theta_s} \leftarrow {\theta_s} - \eta \cdot G_{avg}}$
\end{center}
\State Update $VU_i$ local model parameters:
\begin{center}
    $\theta_i \leftarrow \alpha \cdot \theta_s + (1 - \alpha) \cdot \theta_{avg,i}$
\end{center}
\EndFor
 \State \Return {$ \theta_i, \theta_s$}
\end{algorithmic}
\end{algorithm}

Algorithm 1 details the FSTL process for the considered vehicular scenario. The procedure begins by initializing the VU local model parameters and server model parameters, namely ${\theta_i}$ and ${\theta_s}$, with pre-trained parameters of the DNN used (Alexnet), i.e. ${\theta_i(0)}$ and ${\theta_s(0)}$, respectively, with $i$ and $N$ being the VU index and total number of VUs, and $1\le i\le N$ (lines 1-3). Then, each VU simultaneously performs forward propagation on its local model using its local data. The output of the forward propagation at VU $i$ is denoted as $H_i$, which will then be sent to the server for further forward propagation (lines 4-5). The SL Server on the HAP conducts forward propagation on its server model using this smashed data received from each VU (line 6).  Lines 7 and 8 describe how the SL server calculates the gradients of the server model parameters $\theta_s$ and sends the gradients, ${\nabla\theta_s}$, back to the appropriate VU for additional backpropagation. The local model parameters are then updated using an optimizer, such as stochastic gradient descent (line 9). Here $\eta$ is the learning rate. Next, the FL Server aggregates the gradients received from all VUs by performing federated weighted averaging. Let $w_i$ denote the weight associated with VU $i$ and define $G_i$ as the gradients ${\nabla\theta_s}$ received from VU $i$ (line 10). Afterward, the server model parameters are updated using the aggregated gradients. The SL Server sends back the updated server model parameters ${\theta_s}$ to all participating VUs to update their local model parameters by performing federated averaging (line 11). Here ${\theta_{avg,i}}$ represents the federated average of local updates at VU $i$. Finally, the local model parameters are updated (line 12), where $\alpha$ is a hyper-parameter controlling the weightage of the server model. 

\begin{figure}[tbp]
\centering
\includegraphics[width=\columnwidth]{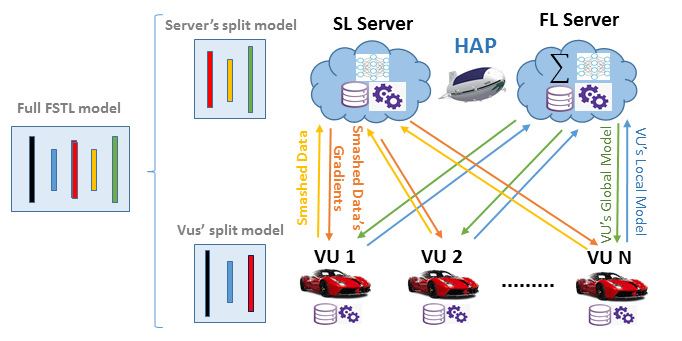}
\caption{FSTL structure for NTNs using HAPs}
\label{Fig:HAP} 
\end{figure}

In the next section, to assess the efficacy of our suggested architecture, we initiated a non-terrestrial vehicular network that is comprised of 20 VUs and one server, acting as both SL server and FL server, which lies on a HAP, as illustrated in Fig. \ref{Fig:HAP}.

\section{Simulations and Performance Evaluations}
The proposed FSTL method, along with the NTN-based vehicular scenario, is simulated over a Python-based platform with the help of additional ML libraries, including Pandas, Numpy, Mathplotlib, etc. In addition, the NVIDIA® Tesla® T4-based GPU accelerator is also used for reduced training intervals.

\begin{figure}[tbp]
\centering
\includegraphics[width=0.91\columnwidth]{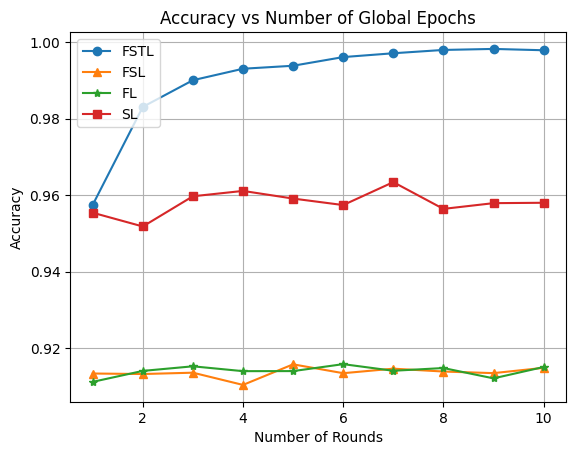}
\caption{Accuracy of FSTL, FSL, FL and SL vs the number of rounds}
\label{Fig:FSTL_accuracy_rounds} 
\end{figure}

The training has been conducted with AlexNet on the MNIST dataset, which contains 60,000 training images and 10,000 testing images. In order to compare the convergence rate of the DL methods, the accuracy of FSTL versus the number of rounds of training has been first evaluated (Fig.~\ref{Fig:FSTL_accuracy_rounds}). A notable observation is that FSTL exhibited a much more rapid increase in accuracy compared to both the original FL and SL. This can be attributed to the employment of knowledge transfer, which is facilitated through the use of TL. Since TL allows leveraging pre-trained models, which have already learned useful features from large-scale datasets, by initializing the client models with pre-trained weights, FSTL can lead to improved model performance compared to starting from scratch in SL or FL. In this way, we can expect FSTL to be able to handle scenarios with heterogeneous data across client devices. The pre-trained model's knowledge provides a robust starting point for all client devices, even if their local datasets vary in size or quality. This helps address challenges associated with data heterogeneity.

\begin{figure}[tbp]
\centering
\includegraphics[width=0.91\columnwidth]{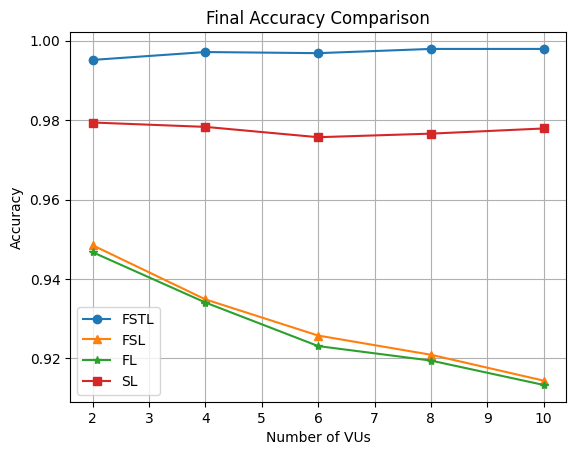}
\caption{Accuracy of FSTL, FSL, FL and SL vs the number of VUs}
\label{Fig:FSTL_accuracy_users} 
\end{figure}

In scenarios where VUs have limited or heterogeneous local data, FSTL helps overcome the challenges of data scarcity. By leveraging the knowledge from a pre-trained model, the client models can benefit from the available local data and achieve good performance even with smaller datasets. In Fig.~\ref{Fig:FSTL_accuracy_users}, we have compared our proposed architecture's final testing accuracy, after 10 global epochs, with that of FL, SL, and FSL. For this purpose, we split the dataset between users, so we decreased the amount of data available for local training. It is evident that in FL, the accuracy drops with an increase in the number of users. On the contrary, FSTL is less sensitive to the number of devices and the amount of data at hand, as we observe almost the same accuracy even with different numbers of VUs. This is a significant advantage of FSTL over other methods.

\begin{figure}[tbp]
\centering
\includegraphics[width=0.91\columnwidth]{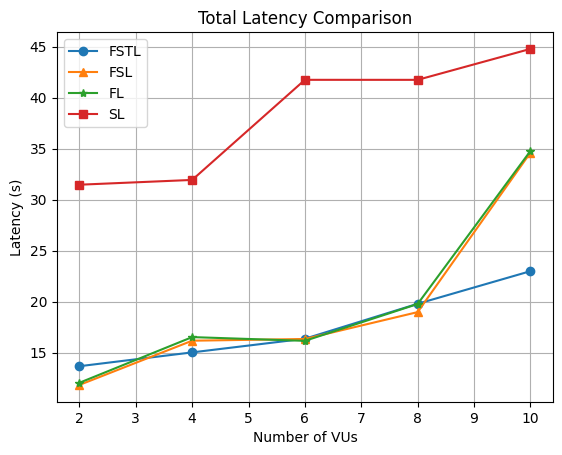}
\caption{Overall Latency for FSTL, FSL, FL and SL vs the number of VUs}
\label{Fig:FSTL_latency} 
\end{figure}

FSTL also reduces communication overhead compared to traditional FL. Instead of sending raw data or gradients, only intermediate representations (derived from SL) are communicated between client devices and the server. This reduces bandwidth requirements and speeds up the training process. To demonstrate this, we conducted a simulation to demonstrate the superior performance of FSTL compared to other methods regarding latency, as the number of users increases. From Fig.~\ref{Fig:FSTL_latency}, it is evident that with higher numbers of VUs, the latency, which is the sum of computation time and communication delay, is significantly higher in SL. The reason for this is that the training process in SL is serial, unlike the parallel aggregation in FL and FSTL, making it much slower compared to the other methods. Additionally, our proposed architecture only trains a portion of the model on resource-constrained devices and communicates solely the gradients of the final (cut) layer on the user side, which is less than the parameters in FL. Therefore, according to Shannon's formula, the communication delay, which is proportional to the data, is reduced, and the processing time in FSTL is minimal compared to other methods. In conclusion, our study demonstrates the effectiveness of FSTL in mitigating latency issues in FSL when the number of users increases.

Although FSTL can enable efficient and accurate ML models over resource-constrained vehicular nodes supported by different NTN platforms, challenges, such as task compatibility and domain shift, which requires a proper pre-trained model aligned to the target/task domain, must be addressed.
%Although FSTL is a promising approach when combined with NTNs in vehicular scenarios where there is a need for efficient and accurate ML algorithms, especially for resource-constrained devices, since exhibits a notably low sensitivity to the number of devices; however, there are challenges to be considered, including task compatibility and domain shift, where selecting an appropriate pre-trained model that aligns with the target task and domain is crucial. 
The transference of the source model's knowledge could be impeded by significant disparities between the source and target data domains, necessitating domain adaptation methods or fine-tuning. Another challenge is model drift, whereby the knowledge of the pre-trained model may not fully align with the local data on client devices. As training progresses, variations in local datasets and optimization dynamics can cause the models to deviate from the pre-trained model, leading to model drift that must be mitigated to maintain optimal performance.

\section{Conclusion}
% This paper introduces a novel method of distributed learning called Transfer Federated Split Learning for vehicular networking scenarios. Overall, TFSL offers the benefits of improved model performance, efficient communication, privacy preservation, and handling data heterogeneity. However, challenges related to compatibility, domain shift, and model drift need to be carefully addressed for successful implementation and deployment.

In this work, we have presented a novel DL approach called FSTL for enabling an efficient FL process by integrating SL and TL while leveraging NTN-based HAPs as a framework for intelligent vehicular networks in the upcoming era of 6G-enabled ITS. Our proposed FSTL over HAPs approach overcomes the limitations of traditional DL methods and offers significant advantages in terms of learning efficiency, accuracy, privacy preservation, and total latency. We have also demonstrated through both latency analysis and the simulations that with a change in the number of DL participants, i.e., VUs, with the same amount of data, the performance in terms of accuracy and latency improves.
%of proposed FSTL method remains constant while traditional approaches such as FL suffers from reduced performance. %Firstly, by leveraging pre-trained models and tailoring them to particular vehicular scenarios, the inclusion of TL improves learning efficiency and accuracy and improves model performance. Secondly, the adoption of SL enables efficient communication by offloading computationally intensive tasks to HAPs, reducing data transmission and preserving privacy through localized model updates. Lastly, the incorporation of FL addresses the challenges of data heterogeneity by enabling collaborative training among VUs, promoting knowledge sharing and adaptation to diverse vehicular scenarios. 
While challenges such as compatibility, domain shift, and model drift require further attention, our framework sets the stage for intelligent VNs by offering enhanced connectivity, privacy preservation, and improved decision-making capabilities. Through integrating advanced DL techniques and utilizing HAPs, our proposed framework holds promise for paving the way for a future of intelligent and connected vehicular systems in NTN environments.

\bibliographystyle{IEEEtran}
\bibliography{IEEEabrv,EW2023}

\end{document}